\documentclass[fleqn,10pt]{wlscirep}
\usepackage[T1]{fontenc}
\usepackage{color}
\usepackage{ulem}

\title{Exactly solvable model of two trapped quantum particles interacting via finite-range  soft-core interactions}
 
\author[1]{Przemys\l aw Ko\'scik}
\author[2,*]{Tomasz Sowi\'nski}
\affil[1]{Institute of Physics, Jan Kochanowski University, ul. \'Swi\c{e}tokrzyska 15, PL-25406 Kielce, Poland}
\affil[2]{Institute of Physics, Polish Academy of Sciences, Aleja Lotnikow 32/46, PL-02668 Warsaw, Poland}
\affil[*]{tomsow@ifpan.edu.pl}

\begin{abstract}
The exactly solvable model of two indistinguishable quantum particles (bosons or fermions) confined in a one-dimensional harmonic trap and interacting via finite-range soft-core interaction is presented and many properties of the system are examined.  Particularly, it is shown that independently on the potential range, in the strong interaction limit bosonic and fermionic solutions become degenerate. For sufficiently large ranges a specific crystallization appears in the system. The results are compared to predictions of the celebrated Busch {\it et al.} model and those obtained in the Tonks-Girardeau limit. The assumed inter-particle potential is very similar to the potential between ultra-cold dressed Rydberg atoms. Therefore, the model can be examined experimentally. 
\end{abstract}
\begin{document}

\flushbottom
\maketitle
\thispagestyle{empty} 

\section*{Introduction}
Exactly solvable Busch {\it et al.} model\cite{Busch} of two ultra-cold bosons confined in a harmonic trap and interacting via contact forces was one of the milestones bringing us closer to our understanding of strongly correlated many-body systems. Although the model deals with only two particles, because of its exact solutions, it inspired many theoretical and experimental studies on collective properties of ultra-cold atoms which are far beyond a simple perturbative description \cite{Blume,Mack,Idziaszek,Astrakharchik,Werner,Idziaszek2,Stetcu,Duerr,Sowinski,Rontani,Sowinski2,March}. Particularly, exact solutions of the model were essential for studies of the Tonks-Girardeau limit of infinite repulsions between particles \cite{Girardeau,Girardeau2,Girardeau3,Kinoshita,Paredes,Murphy,Goold,Lapeyre,Yin}. The validity of the model in a wide range of interactions was finally confirmed in beautiful experiments \cite{Stoferle,Zurn} with a few ultra-cold particles.

Here, we present a wide generalization of the one-dimensional Busch {\it et al.} model to the case of two quantum particles (bosons as well as fermions) interacting via the force of a finite range. With this model and its analytical solutions, it is possible to examine easily different effects caused by the strength and the range of the mutual forces,
restoring well-known results in limiting cases and discovering unsuspected properties in the intermediate regime where the system smoothly transitions between them. Although the model studied seems to be artificial, it approximates the real inter-particle interaction much closer than an oversimplified zero-range potential. Moreover, a proposed shape of the interaction potential can be quite well engineered experimentally in systems of ultra-cold atoms where mutual interactions and shape of an external potential may be controlled with amazing accuracy \cite{Porras,Kim,Carr,Saffman,Islam}. Particularly in the context of ultra-cold Rydberg atoms, due to the Rydberg blockade phenomena, the shape of the inter-particle interaction potential is very close to the shape studied here\cite{Ryd1,Ryd2,Ryd3,Ryd4}. From this point of view, the model studied may have some importance for  understanding of the general problem of two interacting quantum particles.
 
\section{The Model}
In the following we study properties of the system of two identical quantum particles of mass $m$ confined in a one-dimensional harmonic trap of frequency $\Omega$ and interacting via soft-core finite-range rectangular potential. The Hamiltonian of the system reads
\begin{equation} \label{Ham}
{\cal H} = -\frac{\hbar^2}{2m}\left(\frac{\partial^2}{\partial x_1^2}+\frac{\partial^2}{\partial x_2^2}\right)+\frac{m\Omega^2}{2}\left(x_1^2+x_2^2\right)+{\cal V}(x_1-x_2),
\end{equation}
where the interaction potential ${\cal V}(x)$ has a form
\begin{equation}
{\cal V}(x) = \left\{
\begin{array}{ll}
V, &\textrm{if}\,\,|x|< a\\
 0, & \textrm{if}\,\, |x| \geq a
\end{array} \right.\label{inter},
\end{equation}
{\it i.e.}, the interaction energy is constant and it is non-zero only when the distance between particles is not larger than $a$. Similar model was considered previously in three dimensions \cite{DeurePhD}. Depending on the situation we consider symmetric (for bosons) or antisymmetric (for fermions) wave functions with respect to an exchange of particles' positions
\begin{equation} \label{symetria}
\Psi(x_1,x_2) = \pm \Psi(x_2,x_1).
\end{equation}
It is quite natural that in the limit of $a\rightarrow 0$ with constrain $2aV=\mathrm{const}$ one restores the Busch {\it et al.} model with delta-like contact interaction \cite{Busch}, while in the limit $V\rightarrow\infty$ an extensively studied model of hard spheres is obtained \cite{Girardeau}.

Our aim is to give a straightforward and analytical prescription for the eigenstates of the Hamiltonian \eqref{Ham} as a function of the potential depth $V$ and its range $a$. With these solutions, we examine different properties of a few of the lowest eigenstates in the bosonic and fermionic cases. In particular, we consider different single-particle system characteristics (density profile, momentum distribution) as well as inter-particle correlations reflected in a reduced single-particle density matrix.

\section{The Eigenproblem}
To find eigenstates and corresponding eigenenergies of the Hamiltonian \eqref{Ham} it is very convenient to perform standard transformation to the coordinates of the center-of-mass frame:
\begin{align} \label{NewCoord}
R =\frac{x_1+x_2}{2}, \qquad
r =x_2-x_1.
\end{align}
In these new variables the Hamiltonian \eqref{Ham} can be written as a sum of two independent single-particle Hamiltonians ${\cal H}={\cal H}_R + {\cal H}_r$:
\begin{align}
{\cal H}_R &= -\frac{\hbar^2}{4m}\frac{\mathrm{d}^2}{\mathrm{d} R^2}+{m\Omega^2R^2}
, \label{HamR} \\
{\cal H}_r &= -\frac{\hbar^2}{m}\frac{\mathrm{d}^2}{\mathrm{d} r^2}+\frac{m\Omega^2}{4}r^2+{\cal V}(r). \label{Hamr}
\end{align}
The Hamiltonian ${\cal H}_R$ has a textbook form of the harmonic oscillator and it can be diagonalized straightforwardly. 
The Hamiltonian ${\cal H}_r$ has an additional term related to the interactions \eqref{inter} and the corresponding eigenequation, when written in the natural units of an external harmonic oscillator, has the form:
\begin{equation} \label{EigenEq}
\left[-\frac{\mathrm{d}^2}{\mathrm{d} r^2}+\frac{1}{4}r^2+{\cal V}(r)\right]\Phi(r) = E \Phi(r).
\end{equation}
Since the Hamiltonian ${\cal H}_r$ commutes with the operator of the parity inversion, ${\cal P}\!:r \rightarrow -r$, the eigenstates of ${\cal H}_r$ can be chosen as either even or odd functions of the relative position~$r$. They directly correspond to bosonic and fermionic statistics, respectively.

The eigenequation \eqref{EigenEq} has the form of the Weber differential equation \cite{Weber}:
\begin{equation} \label{WebberEq}
\left(-\frac{\mathrm{d}^2}{\mathrm{d}r^2}+\frac{1}{4} r^2\right) \Phi(r)=-\left(u+\frac{1}{2}\right) \Phi(r),
\end{equation}
with $u$ equal to $-E+V-1/2$ and $-E-1/2$ for $|r|<a$ and  $|r|\geq a$, respectively. The Weber equation was originally studied to solve Laplace equation expressed in parabolic coordinates \cite{Weber} but it appears in different problems of mathematical physics and many of its properties are well known \cite{Merzbacher,Abramowitz}. In the case studied, when the problem is not reduced to the ordinary harmonic oscillator problem ($V\neq 0$), it is very convenient to consider two different pairs of the solutions $\{\varphi_u^{(+)}(r),\varphi_u^{(-)}(r)\}$
and $\{\phi_u^{(+)}(r),\phi_u^{(-)}(r)\}$ having appropriate symmetry under parity transformation ${\cal P}$ but different properties on the boundaries. The first pair can be expressed in terms of the confluent hypergeometric function $_1\mathbf{F}_1$ as:

\begin{align}
\varphi^{(+)}_u(r) &=\mathrm{e}^{-\frac{r^2}{4}} \, _1\mathbf{F}_1\left(\frac{u+1}{2};\frac{1}{2};\frac{r^2}{2}\right), \label{WebberEq1a} \\
\varphi^{(-)}_u(r) &=r\,\mathrm{e}^{-\frac{r^2}{4}} \, _1\mathbf{F}_1\left(\frac{u+2}{2};\frac{3}{2};\frac{r^2}{2}\right). \label{WebberEq1b}
\end{align}
These functions may be consider as appropriate solutions only in the region $|r|< a$ because they are divergent in the infinity, $r\rightarrow \pm\infty$. The second pair of solutions is expressed in terms of other confluent hypergeometric function $\mathbf{U}$:
\begin{align}
\phi^{(+)}_u(r) &= \mathrm{e}^{-\frac{r^2}{4}} \mathbf{U}\left (\frac{u+1}{2};\frac{1}{2};\frac{r^2}{2}\right), \label{WebberEq2a}\\
\phi^{(-)}_u(r) &= r\,\mathrm{e}^{-\frac{r^2}{4}} \mathbf{U}\left(\frac{u+2}{2};\frac{3}{2};\frac{r^2}{2}\right). \label{WebberEq2b}
\end{align}
\begin{figure}
\includegraphics[width=0.7\linewidth]{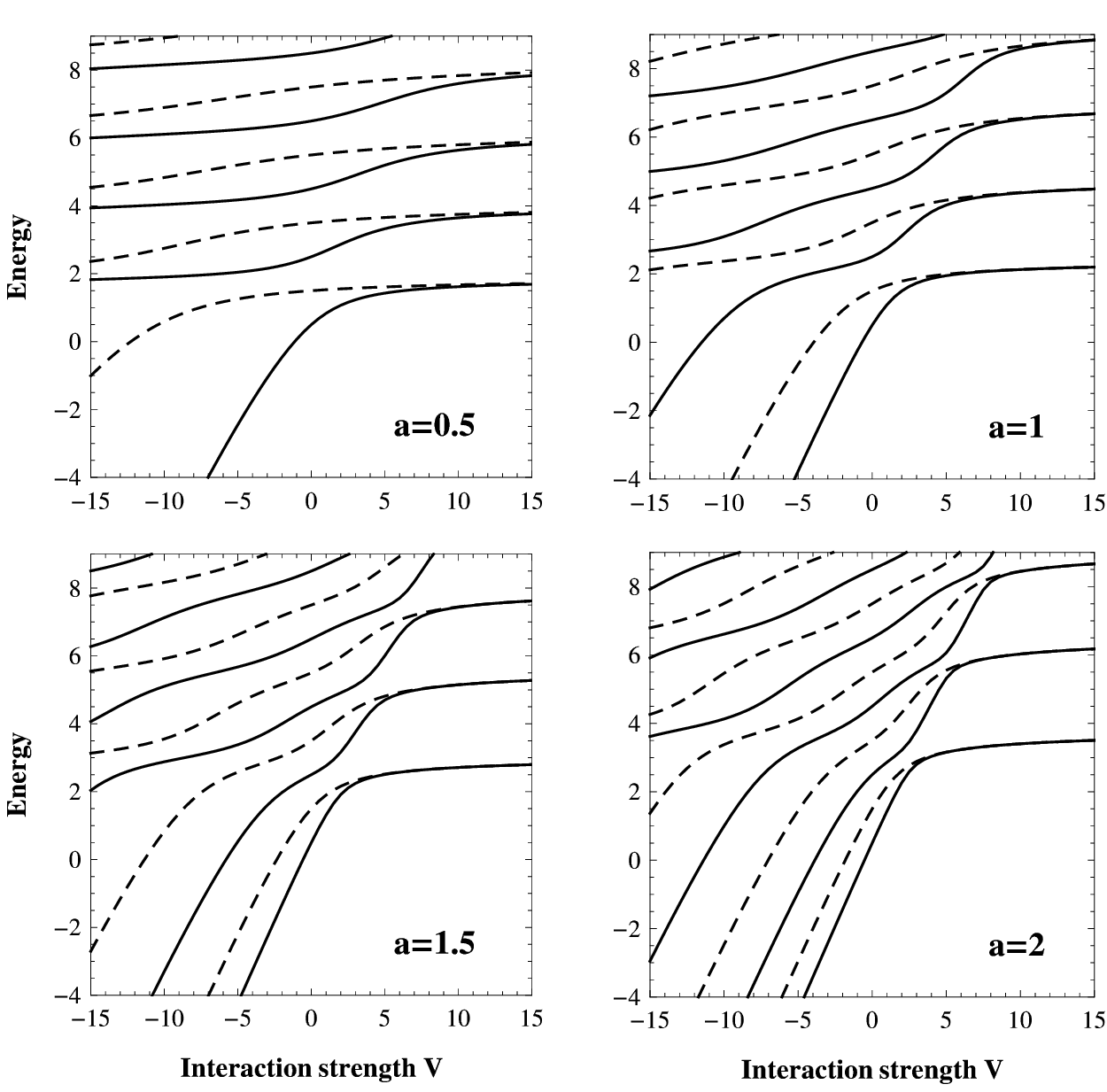}
\caption{Spectrum of the relative motion Hamiltonian ${\cal H}_r$ as a function of interaction strength $V$ and chosen potential range $a$. For clarity, even (odd) solutions corresponding to bosonic (fermionic) cases are plotted with solid (dashed) lines. Note that specific degeneracy between neighboring eigenenergies is established in the hard-core limit ($V\rightarrow\infty$). Energy and interaction strength $V$ are measured in units of $\hbar\Omega$. \label{Fig1}}
\end{figure}
In contrast to the first pair, these functions decay appropriately in the limit $r\rightarrow \pm\infty$ but they do not have appropriate behavior at $r=0$, {\it i.e.}, odd functions $\phi^{(-)}_u(r)$ are discontinuous, whereas even ones $\phi^{(+)}_u(r)$ have discontinuous first derivative. It means that functions $\phi^{(\pm)}_u(r)$ may be considered as appropriate solutions only in the region $|r|\geq a$. Consequently, any solution of the eigenequation \eqref{EigenEq} with energy $E_i$ (which is continuous and has a continuous first derivative in a whole space) may be constructed as following
\begin{equation} \label{solu}
\Phi^{(\pm)}_{i}(r) = {\cal N}_i^{(\pm)}\left\{ \begin{array}{ll}
  A^{(\pm)}_\nu\varphi^{(\pm)}_\nu(r), & |r|<{a} \\
\phi^{(\pm)}_\mu(r), & |r|\geq{a}
\end{array}\right. ,
\end{equation}
where $\nu=-E_i+V-1/2$, $\mu= -E_i-1/2$ and ${\cal N}_i^{(\pm)}$ is a normalization coefficient of the resulting function. An additional coefficient $A^{(\pm)}_\nu$ together with the eigenenergy $E_i$ are determined by matching conditions at $|r|=a$, ensuring that the function \eqref{solu} and their first derivatives are continuous in the whole space:
\begin{align}
A_\nu^{(\pm)}\varphi_\nu^{(\pm)}(a)&=\phi_\mu^{(\pm)}(a), \label{transa}\\
\left.A_\nu^{(\pm)}\frac{\mathrm{d}}{\mathrm{d}r}\varphi_\nu^{(\pm)}(r)\right|_{r=a} &= \left.\frac{\mathrm{d}}{\mathrm{d}r}\phi_\mu^{(\pm)}(r)\right|_{r=a}. \label{transb}
\end{align}
As typical for such problems, the conditions \eqref{transa} and \eqref{transb} can be fulfilled only for some particular values of an eigenenergy $E_i$ leading directly to the quantization of the physical spectrum. In the case studied, the matching conditions \eqref{transa} and \eqref{transb} are fulfilled when eigenenergy $E_i$ is a solution of the following transcendental equation
\begin{equation}
\phi_\mu^{(\pm)}(a)\left.\frac{\mathrm{d}}{\mathrm{d}r}\varphi_\nu^{(\pm)}(r)\right|_{r=a}-
\varphi_\nu^{(\pm)}(a)\left.\frac{\mathrm{d}}{\mathrm{d}r}\phi_\mu^{(\pm)}(r)\right|_{r=a}=0, \nonumber
\end{equation}
where $\nu=-E_{i}+V-1/2$, $\mu= -E_{i}-1/2$. It directly leads to the following equations
\begin{equation} \label{App1}
2 (\nu +1) \, _1\mathbf{F}_1\left(\frac{\nu +3}{2};\frac{3}{2};\frac{a^2}{2}\right) \mathbf{U}\left(\frac{\mu +1}{2};\frac{1}{2};\frac{a^2}{2}\right)+(\mu +1) \,
   _1\mathbf{F}_1\left(\frac{\nu +1}{2};\frac{1}{2};\frac{a^2}{2}\right) \mathbf{U}\left(\frac{\mu +3}{2};\frac{3}{2};\frac{a^2}{2}\right)=0,
\end{equation}
and
\begin{equation} \label{App2}
3 (\mu +2) \, _1\mathbf{F}_1\left(\frac{\nu +2}{2};\frac{3}{2};\frac{a^2}{2}\right) \mathbf{U}\left(\frac{\mu +4}{2};\frac{5}{2};\frac{a^2}{2}\right)+2 (\nu +2) \,
   _1\mathbf{F}_1\left(\frac{\nu +4}{2};\frac{5}{2};\frac{a^2}{2}\right) \mathbf{U}\left(\frac{\mu +2}{2};\frac{3}{2};\frac{a^2}{2}\right)=0
\end{equation}
determining even and odd solutions, respectively.
Equations \eqref{App1} and \eqref{App2} are quite complicated but they can be solved straightforwardly with simple numerical methods. After determining eigenenergies one finds the corresponding  coefficients
\begin{equation}
A_{\nu}^{(\pm)}=\frac{\phi_{\mu}^{(\pm)}(a)}{\varphi_{\nu}^{(\pm)}(a)},
\end{equation}
and thus corresponding wave functions of the relative motion \eqref{solu}.

It is worth mentioning that in the limiting case of noninteracting particles ($V=0$) standard solutions of a one-dimensional harmonic oscillator are restored. Indeed, in this particular case one finds $\mu=\nu=-E_i -1/2$ and the matching conditions \eqref{transa} and \eqref{transb} reduce to a simple demanding that eigenenergies are half-integer numbers, $E_i=i + 1/2$. In consequence, appropriate functions $\varphi^{(\pm)}_\nu(r)$ and $\phi^{(\pm)}_\mu(r)$ become equivalent and they are expressed in terms of Hermite polynomials $\mathbf{H}_i$:
\begin{equation}
\Phi_i(r) = {\cal N}_i\,\mathrm{e}^{-\frac{r^2}{4}}\,\mathbf{H}_i\left(\frac{r}{\sqrt{2}}\right).
\end{equation}
In the opposite limit of infinite repulsions ($V\rightarrow\infty$) situation is also simplified. In this case the relative wave functions \eqref{solu} must vanish in a whole range $|r|<a$, {\it i.e.}, all amplitudes $A_\nu^{(\pm)}=0$. It immediately leads to the simplified quantization condition
\begin{equation}
\phi^{(\pm)}_\mu(a)=0,
\end{equation}
and in consequence to the typical for hard-core limit degeneracy between neighboring even and odd solutions.

Having analytical solutions \eqref{solu} one can express any eigenstate of the Hamiltonian \eqref{Ham} as a simple product of two wave functions
\begin{equation} \label{EigenGeneral}
\Psi_{ij}(x_1,x_2) = \Upsilon_i\left(\frac{x_1+x_2}{2}\right)\Phi_j(x_1-x_2),
\end{equation}
where $\Upsilon_i(R)$ and $\Phi_j(r)$ are appropriate eigenstates of the center-of-mass and relative motion Hamiltonians, respectively. Although these two particular coordinates ($R$ and $r$) are completely decoupled, the wave function \eqref{EigenGeneral} cannot be written (for any finite $V$) as a product (for bosons) or single Slater determinant (for fermions) of wave functions of independent particles. This observation leads directly to non trivial quantum correlations between particles which are discussed in the following.

\section*{Spectral properties}
Many properties of the system studied can be extracted directly from the spectrum of the relative motion Hamiltonian \eqref{Hamr}. In Fig.~\ref{Fig1} we show several the lowest eigenenergies of ${\cal H}_r$ as functions of the interaction strength $V$ for different potential ranges $a$. Solid lines correspond to even wave functions (bosons) while dashed lines to odd cases (fermions). As suspected, for $V=0$ the spectrum of noninteracting particles is restored, {\it i.e.}, alternating bosonic (symmetric) and fermionic (antisymmetric) states of the relative motion have equally distributed energies $\hbar/2$, $3\hbar/2$, $5\hbar/2$, {\it etc}. For non-vanishing interactions $V\neq 0$, depending on the potential range $a$, eigenenergies vary. A rapidity of these changes crucially depends on a sign of interactions -- for the attraction is much higher than for the repulsion. It is interesting to note that for any finite range ($a\neq 0$) and sufficiently large attractions each eigenstate may have arbitrary large negative energy.  Note also that, independently on the potential range $a$, energies never cross. It means that for any finite $a$ and $V$ any eigenstate of the system is not degenerated. In fact, it is a direct consequence of one-dimensionality of the relative motion Hamiltonian ${\cal H}_r$ \cite{LandauBook}.

In the particular limit of strong repulsions, the neighboring states of opposite symmetry become degenerate independently on the interaction range $a$. This observation is a direct consequence of the Bose-Fermi mapping \cite{Girardeau,Girardeau2}.

\begin{figure}
\includegraphics[width=0.5\linewidth]{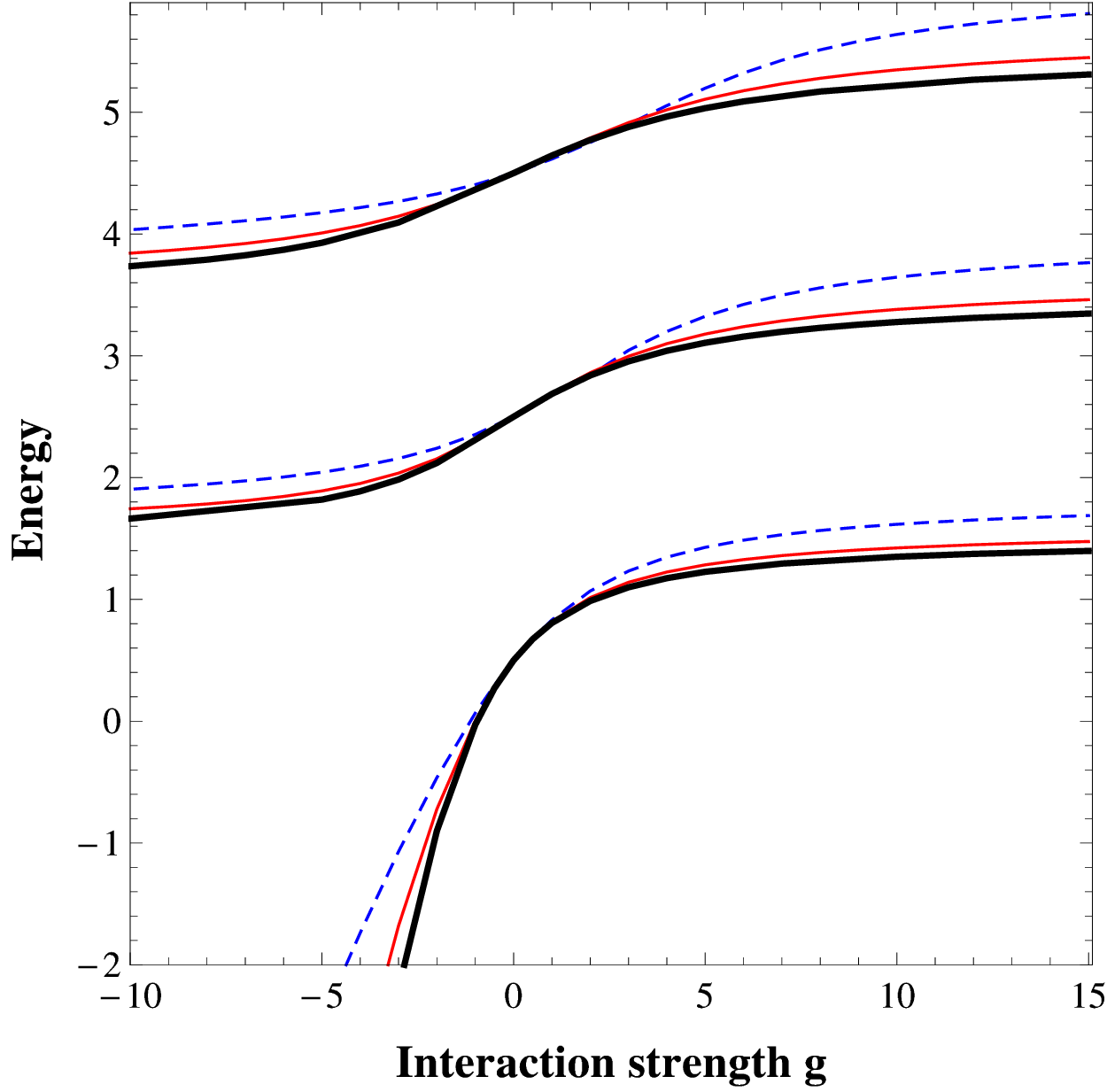}
\caption{Even (bosonic) part of the spectrum of the relative motion Hamiltonian ${\cal H}_r$ as a function of rescaled interaction strength $g=2aV$ for two different values of the interaction range: $a=0.5$ (thin dashed  blue line) and $a=0.15$ (thin solid red line). Along with vanishing $a$ the limiting case of contact forces $a\rightarrow 0$ (thick black line) is obtained in a wide range of interactions. Energy and rescaled interaction strength $g$ are measured in units of $\hbar\Omega$ and $\sqrt{\hbar^3\Omega/m}$, respectively. \label{Fig2}}
\end{figure}

It is clearly seen in Fig.~\ref{Fig1} that properties of even and odd solutions of the relative coordinate eigenproblem are essentially different when the potential range $a$ becomes smaller than the natural length scale of the problem (in dimensionless units equal to $1$). As it is seen, in contrast to bosonic states, energies of fermionic states (dashed lines) become more horizontal, {\it i.e.}, they are less sensitive to the interaction energy strength. This behavior is a consequence of the fact that odd functions always vanish at $r=0$, {\it i.e.}, along with decreasing $a$ the interaction energy rapidly decreases independently on interaction strength $V$. In the limit of vanishing $a$ the interaction is completely described in terms of the $s$-wave scattering which is not present between indistinguishable fermionic particles.
\begin{figure}
\includegraphics[width=0.8\linewidth]{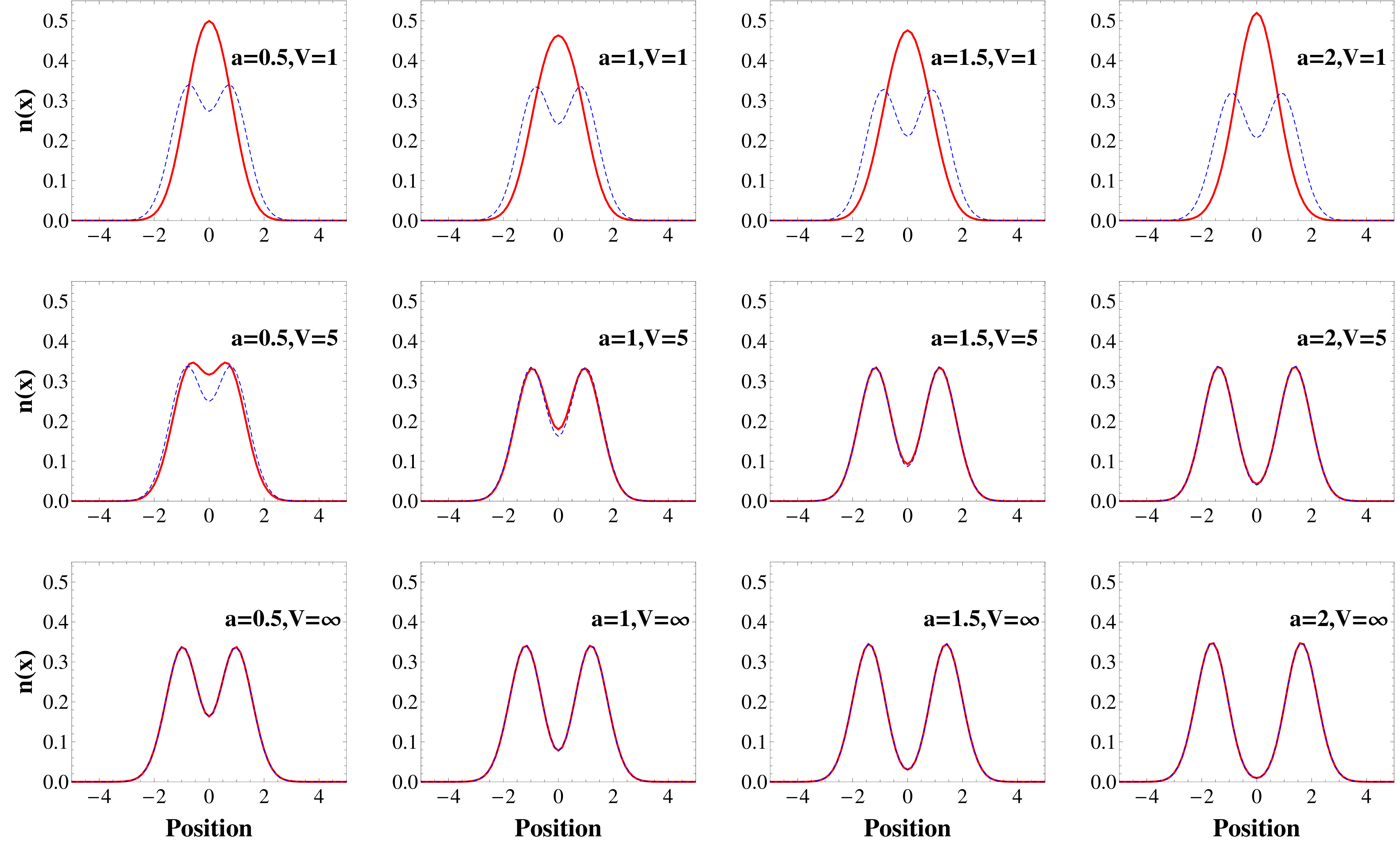}
\caption{The density distribution \eqref{DensD} calculated for bosonic (solid red line) and fermionic (dashed blue line) ground-states for different ranges of the potential $a$ (columns) and different potential strengths $V$ (rows). Note that for a sufficiently large strength $V$ the density profile is the same for both statistics (see main text for details). The positions and the densities are measured in units of
$\sqrt{\hbar/(m\Omega)}$ and $\sqrt{m\Omega/\hbar}$, respectively.\label{Fig3}}
\end{figure}

\section*{Contact interactions limit}
Mentioned above limiting case of contact forces can be explored more precisely by considering a formal limit in which the interactions ${\cal V}(r)$ become identical with $\delta$-like potential of the form $g\delta(r)$, {\it i.e.}, in the limit $a\rightarrow 0$ with fixed product $2aV=g=\mathrm{const}$. In this limit, the problem reduces to the celebrated model of two quantum particles interacting via contact forces for which exact analytical solutions are known \cite{Busch}. To show how the limiting spectrum is restored we fix the potential range $a$ and for given limiting interaction $g$ we calculate a rescaled value of potential strength $V=g/(2a)$. In this way one obtains the spectrum of the relative motion Hamiltonian ${\cal H}_r$ for different ranges $a$ rescaled to the interaction strength $g$ of the Busch {\it et al.} model. Results of this procedure adopted to even (bosonic) eigenstates of the relative motion Hamiltonian are presented in Fig.~\ref{Fig2}. As it is seen, with decreasing potential range $a$ corresponding eigenenergies approach the results for contact interactions (thick black line). For $a=0.15$ (red solid line) an agreement is almost perfect in a wide range of interactions. These results are in qualitative agreement with previously obtained finite range corrections obtained within the Green's function approach for higher dimensionality \cite{Zinner}.

At this point, it should be noted that for any finite $a$, in contrast to the contact interactions limit, all eigenenergies become negative for sufficiently strong attractions (see Fig.~\ref{Fig1}). Only in the case of contact interactions there exists exactly one bound state of the Hamiltonian \eqref{Hamr} -- the ground-state of the bosonic system.

\section*{Single-particle quantities}
The simplest quantities which can be measured experimentally quite easily are related to single-particle properties. All of them are fully captured by the reduced single-particle density matrix which for the model studied has a form:
\begin{equation}
\Gamma(x,x') = \int_{-\infty}^\infty\!\mathrm{d}x_2\, \Psi^*(x,x_2)\Psi(x',x_2),
\end{equation}
where $\Psi(x_1,x_2)$ is a chosen two-particle state of the system. In the following we focus on the properties of the bosonic and the fermionic ground-states, {\it i.e.}, according to the notation of eq. \eqref{EigenGeneral} the states $\Psi_{00}(x_1,x_2)$ and $\Psi_{01}(x_1,x_2)$, respectively.

\begin{figure}
\includegraphics[width=0.8\linewidth]{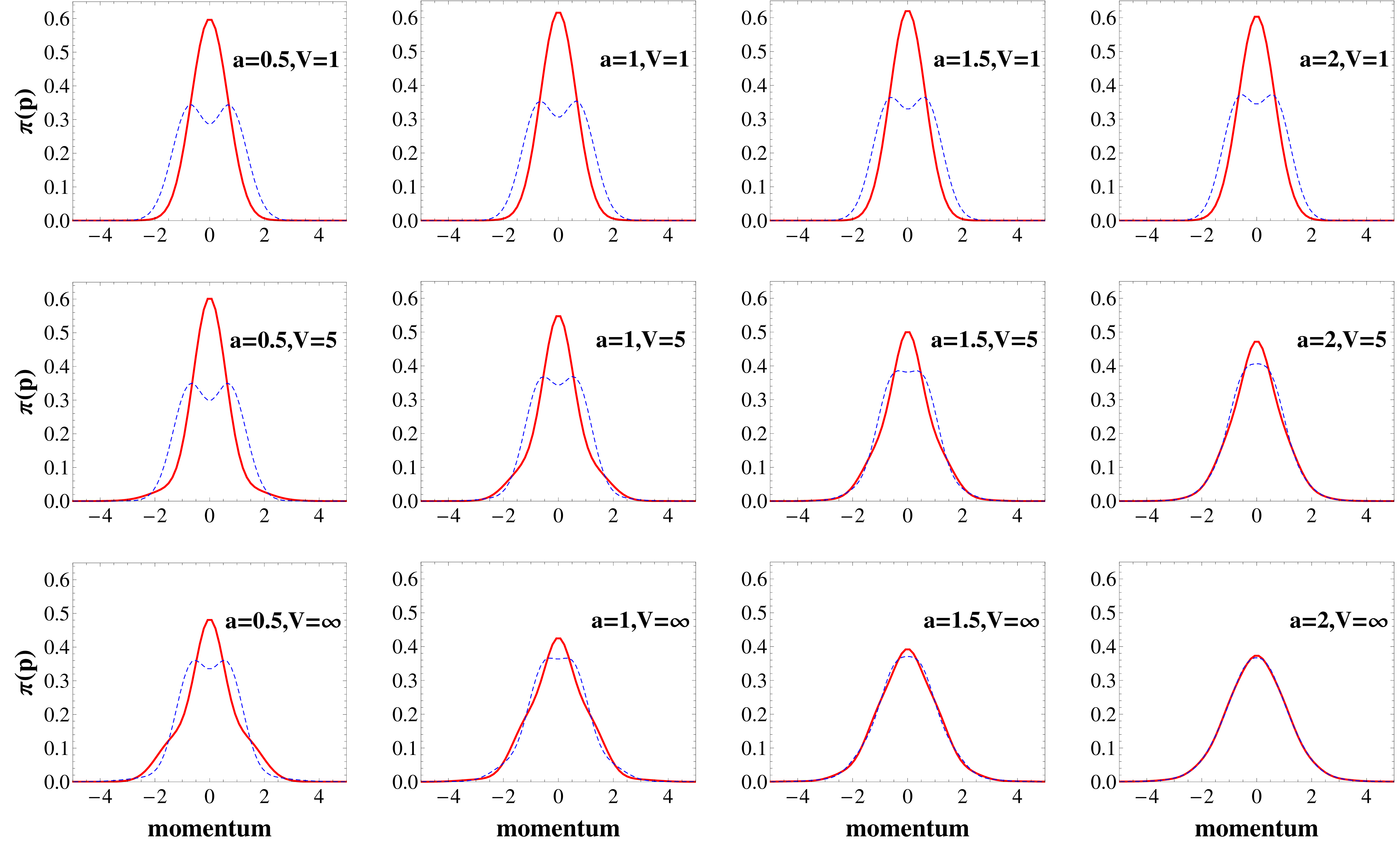}
\caption{The distribution of the single-particle momentum \eqref{MomentumD} calculated for bosonic (solid red line) and fermionic (dashed blue line) ground-states for different ranges of the potential $a$ (columns) and different potential strengths $V$ (rows). Although, for sufficiently large strength $V$ density profiles in Fig.~\ref{Fig3} are the same for both statistics, momentum distribution not necessarily have this property. Only for large enough ranges both distributions become equal indicating appearance of crystallization (see main text for details). The momentum and the momentum distributions are measured in units of $\sqrt{m\hbar\Omega}$ and $\sqrt{1/(m\hbar\Omega)}$, respectively.\label{Fig4}}
\end{figure}
Typically, we are mostly interested not in the whole reduced single-particle density matrix $\Gamma(x,x')$ but only in its diagonal part
\begin{equation} \label{DensD}
n(x)=\Gamma(x,x),
\end{equation}
which represents a density profile of particles. Analogously, a diagonal part of its Fourier transform
\begin{equation} \label{MomentumD}
\pi(p) = {1\over 2\pi}\int_{-\infty}^\infty\mathrm{d}x\int_{-\infty}^\infty\mathrm{d}x'\,\, \Gamma(x,x')\,\mathrm{e}^{i p(x-x')/\hbar},
\end{equation}
encodes distribution of a single-particle momentum. Properties of these two simple quantities crucially depend on the range of the potential $a$. These differences are especially manifested in the cases which are beyond applicability of the Busch {\it et al.} model.
In Fig.~\ref{Fig3} and Fig.~\ref{Fig4} we show density and momentum distributions for bosonic (red solid line) and fermionic (dashed blue line) ground-states obtained for a few representative potential ranges ($a=1$, $a=1.5$, and $a=2$) and different potential strengths $V$, including hard-core limit case $V\rightarrow\infty$.

Let us recall that in the hard-core limit, the bosonic wave functions necessarily satisfy the condition $\Psi_{00}(x_{1},x_{2})=0$ on the line $x_{1}=x_{2}$, regardless of $a$ since corresponding wave functions of relative motion $\Phi(r)$ vanish at $r=0$. This observation, usually called fermionization, enables one to map the bosonic ground-state wave function to the fermionic one via the following relation $\Psi_{00}(x_{1},x_{2})=|\Psi_{01}(x_{1},x_{2})|$. In consequence, in the hard-core limit, the bosonic and fermionic ground states share not only the same energy but also have the same spatial density profiles (bottom row in Fig.~\ref{Fig3}). Note however, that this particular mapping (forced by infinite repulsions) does not necessarily mean that also momentum distributions for bosonic and fermionic ground-states are the same (bottom row in Fig.~\ref{Fig4}).

\begin{figure}
\includegraphics[width=0.5\linewidth]{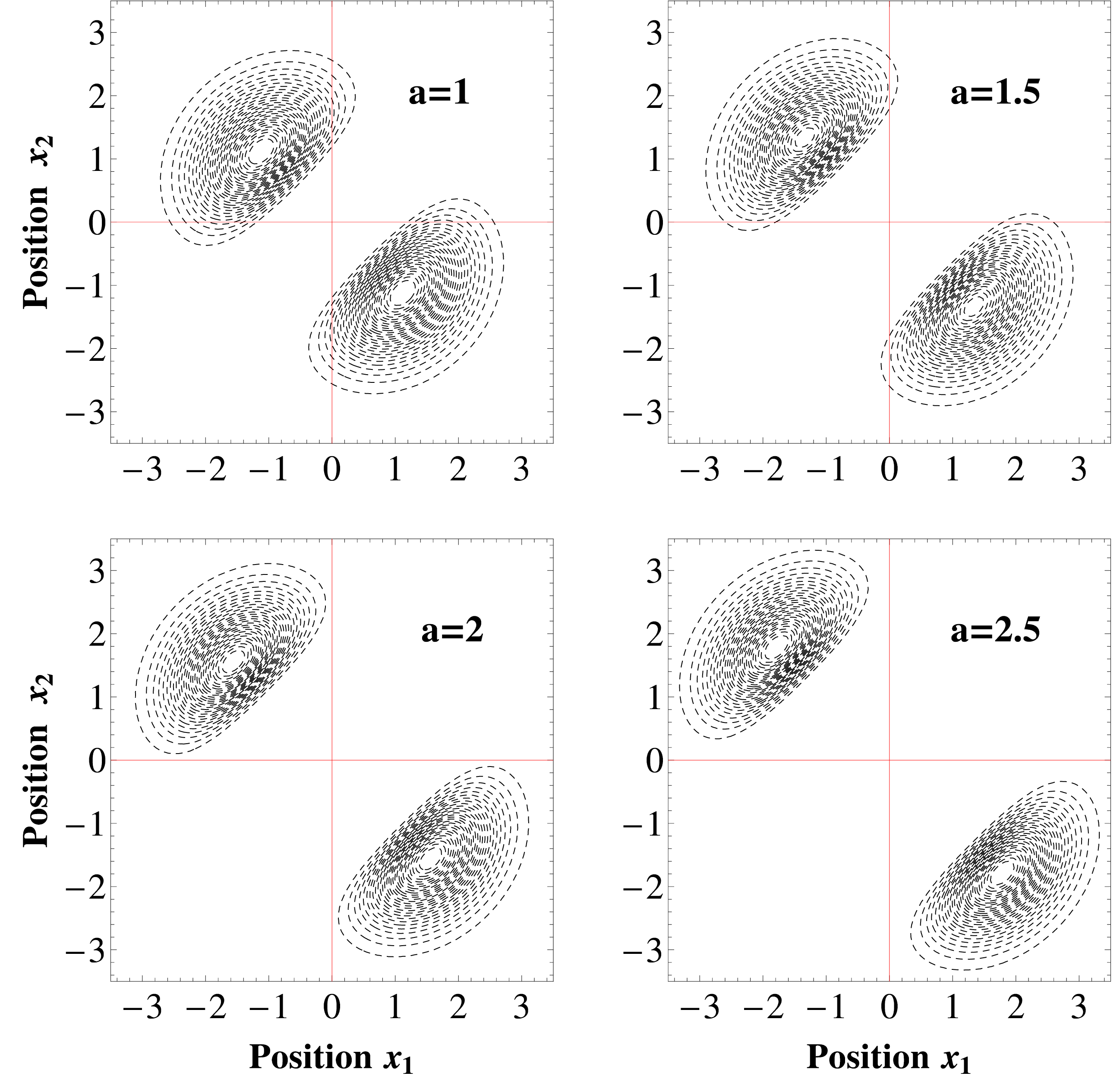}
\caption{Two-particle probability density of finding particles at positions $x_1$ and $x_2$ in hard-core limit $V\rightarrow\infty$ for different values of the potential range $a$. For sufficiently large range ($a\gtrsim 2$) particles occupy exactly opposite sides of the trap. This is one of the features of the crystallization mechanism. The positions and the probability distributions are measured in units of $\sqrt{\hbar/(m\Omega)}$ and $m\Omega/\hbar$, respectively.\label{Fig5} }
\end{figure}

The situation becomes essentially different for potential ranges $a\gtrsim 2$. In these cases, not only the density profiles $n(x)$ but also the momentum distributions $\pi(p)$ of bosonic and fermionic ground-states become identical in the hard-core limit (right bottom plots in Figs.~\ref{Fig3} and \ref{Fig4}). At the same time the density profiles exhibit spatial separation into two independent peaks indicating localization of particles at opposite sides of the trap. This result is clearly understandable when two-particle density profile $\rho(x_1,x_2)=|\Psi(x_1,x_2)|^2$ is considered. As it is seen in Fig.~\ref{Fig5} the probability of finding both particles at the same side of the trap ($x_1,x_2>0$ or $x_1,x_2<0$) vanishes when $a\gtrsim 2$. It is often said that the system enters the crystallization regime where individual particles are spatially separated\cite{Deure} and behave as distinguishable parties. In consequence, any physical property of the system does not depend on a quantum statistics. However, as explained in the next section, spatial separation does not mean that individual particles can be treated as independent since non-classical correlations between them are still present.

For finite interaction strengths $V<\infty$ (see Figs.~\ref{Fig3} and \ref{Fig4}) one can observe how the quasi-fermionization is built along with increasing $V$. It is quite instructive to note that the fermionization regime is reached earlier if the interaction range $a$ is bigger. It is consistent with previous results concerning quasi-degeneracy in the spectrum of the relative motion Hamiltonian (compare to Fig.~\ref{Fig1}).

\section*{Inter-particle correlations}
As was mentioned before, the eigenstates \eqref{EigenGeneral} are always separable when written with respect to coordinates \eqref{NewCoord}. However, for any finite interaction, it cannot be written as a product with respect to positions of particles $x_1$ and $x_2$. This simple observation means that the eigenstates of the interacting system studied encode nonclassical correlations between particles. These inter-particle correlations are nicely captured by spectral properties of the reduced single-particle density matrix $\Gamma(x,x')$. It is known that the matrix $\Gamma(x,x')$ can be decomposed to its natural Schmidt orbitals
\begin{equation} \label{sd}
\Gamma(x,x') =\sum_{i}\lambda_{i} u_{i}(x)u_{i}(x'),
\end{equation}
where $u_{i}(x)$ and $\lambda_{i}$ are eigenvectors and eigenvalues of the reduced density matrix $\Gamma(x,x')$, respectively. Coefficients $\lambda_i$ have a direct interpretation of probabilities of finding a single particle in quantum states described by the corresponding orbitals and they are normalized to unity, $\sum_i \lambda_i = 1$. Let us note that in the case of fermions, due to Pauli exclusion principle \eqref{symetria}, all non-zero eigenvalues are doubly degenerated.

If both bosons occupy exactly the same orbital $u_0(x)$ then the reduced density matrix simply projects to the orbital $u_0(x)$ and particles are trivially correlated. What is less intuitive, correlations between particles are also trivial whenever particles occupy two different orbitals $u_0(x)$ and $u_1(x)$ in the way that the two-particle wave function is represented by their Slater determinant (for fermions) or permanent (for bosons). In all these cases, the state is regarded as non-entangled since correlations originate only in the quantum statistics of indistinguishable particles \cite{Schliemann,Ghirardi}. In other cases, additional correlations are forced by mutual interactions and they are directly reflected in increasing number of non-vanishing occupations $\lambda_i$ in the decomposition \eqref{sd}.

\begin{figure}
\includegraphics[width=0.9\linewidth]{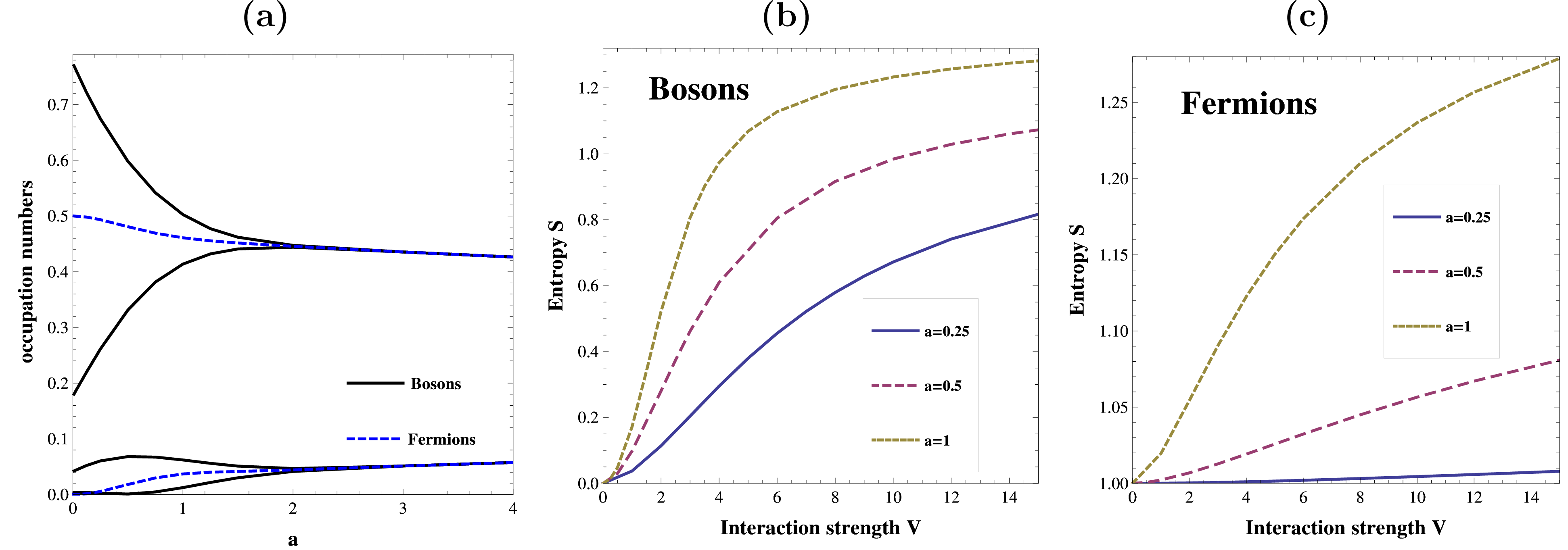}
\caption{(a) First four of the largest eigenvalues of the reduced single-particle density matrix $\Gamma(x,x')$ calculated for the bosonic (solid black lines) and fermionic (dashed blue lines) ground-state in the hard-core limit $V\rightarrow\infty$ as functions of the potential range $a$. In the limit of contact interactions ($a\rightarrow 0$) fermionic ground-state does not manifest any non-trivial correlations. For large ranges ($a\gtrsim 2$) bosonic eigenvalues become degenerate and equal to appropriate fermionic ones indicating crystallization regime. Note that fermionic eigenvalues are always doubly degenerated. (b-c) The von Neumann entropy $\mathbf{S}$ as a function of interaction strength $V$ for different potential ranges $a$. Note that in the fermionic case its growth is strongly suppressed in the limit of vanishing range ($a\rightarrow 0$) as a consequence of vanishing contact interaction in this limit. In all plots the potential range $a$ and strength $V$ are measured in units of $\sqrt{\hbar/(m\Omega)}$ and $\hbar\Omega$ respectively. \label{Fig6}}
\end{figure}

In the Fig.~\ref{Fig6}a we present four the largest eigenvalues $\lambda_i$ of the reduced density matrix $\Gamma(x,x')$ calculated for bosonic (solid black lines) and fermionic (dashed blue line) ground-states, $\Psi_{00}(x_1,x_2)$ and $\Psi_{01}(x_1,x_2)$, as functions of potential range $a$ in the hard-core limit $V\rightarrow\infty$. As noted above, in the fermionic case eigenvalues are doubly degenerated. In the contact interaction limit ($a\rightarrow 0$), fermions, in contrast to bosons,  become noninteracting and therefore there is no additional correlation beyond that induced by quantum statistics ($\lambda_0=\lambda_1=0.5$). In opposite limit of large ranges $a\gtrsim 2$, bosonic occupations $\lambda_i$ become doubly degenerated and equal to fermionic ones. Simultaneously, reduced single-particle matrices of bosonic and fermionic ground-states become identical. By a direct inspection of the two-particle state we found that the ground-state of the system can be written as ($\pm$ sign for bosons and fermions, respectively):
\begin{equation}
\Psi(x_1,x_2) = \sum_i \kappa_i \left[{\cal L}_{i}(x_1){\cal R}_{i}(x_2)\pm {\cal L}_{i}(x_2){\cal R}_{i}(x_1)\right], \nonumber
\end{equation}
where ${\cal L}_i(x)$ and ${\cal R}_i(x)$ are single-particle orbitals localized in left and right side of the trap constructed from the corresponding even and odd single-particle orbitals of the reduced density matrix $\Gamma(x,x')$ \cite{Koscik}:
\begin{align}
{\cal L}_i(x) &= \frac{1}{\sqrt{2}}\left[u_{2i}(x)+u_{2i+1}(x)\right], \\
{\cal R}_i(x) &= \frac{1}{\sqrt{2}}\left[u_{2i}(x)-u_{2i+1}(x)\right].
\end{align}
The construction is possible, since in this case the appropriate eigenorbitals are degenerated and any of their linear combination remains as an eigenvector of $\Gamma(x,x')$. The amplitudes $\kappa_i$ are related directly to the occupations $\lambda_i$, $\kappa_i=\sqrt{\lambda_{2i}}=\sqrt{\lambda_{2i+1}}$.
This observation is one of quite spectacular manifestations of the crystallization mentioned before. Note that the values of the dominant occupations $\lambda_0$ and $\lambda_1$ decreases with potential range $a$. It means that even in the crystallization regime particles cannot be treated as trivially correlated parties and therefore they cannot be locally described with individual well-defined orbitals.

Non-classical correlations between particles are quite well quantified by the von Neumann entropy. When occupancies $\lambda_i$ are known, it can be calculated straightforwardly as ${\mathbf S} = -\sum_i \lambda_i \log_2\lambda_i$. This measure exactly vanishes in the non-entangled  product state of bosons ($\lambda_0=1$) and it is equal to 1 in the non-entangled bosonic and fermionic states ($\lambda_0=\lambda_1=0.5$). Note however that in the bosonic case, an opposite implication does not hold, {\it i.e.}, the condition $\mathbf{S}=1$ does not necessarily means that the state in non-entangled \cite{Ghirardi}. In Fig.~\ref{Fig6}b (for bosons) and Fig.~\ref{Fig6}c (for fermions) we show the dependence of the von Neumann entropy on the potential strength $V$ for different ranges $a$ calculated in the ground-state of interacting particles. It is clear that in a considered range of parameters the von Neumann entropy $\mathbf{S}$ is a monotonic function of the interaction $V$ and its growth crucially depends on the potential range $a$. It is also worth noticing that in the fermionic case and vanishing range ($a\rightarrow 0$) the von Neumann entropy remains unchanged independently on the interaction strength $V$. This is a direct consequence of vanishing contact interaction for fermionic species.

\section*{Summary}
In this paper, we present properties of the exactly solvable model of two interacting particles confined in a harmonic trap. Inter-particle forces are modeled by a square wall controlled by two independent parameters: potential range and its strength. The results enabled us to investigate and discuss different properties of the system in a whole range of parameters between limiting cases of well known Busch {\it et al.} and hard-core models. Obtained results suggest that any finite range of the inter-particle forces is directly reflected in simple quantities which can be measured experimentally. The prominent example is the many-body spectrum where, in contrast to the zero-range case, all eigenstates become unbounded from below for attractive forces. We show that also density and momenta distributions maybe strongly affected by the finite range of the potential. All these deviations from the Busch {\it et al.} model maybe examined in recent experiments on a few ultra-cold particles.

The model presented belongs to the specific class of quite realistic quantum many-body problems having exact analytical solutions \cite{UshveridzeBook,LiebBook,Korepin,SutherlandBook}. From this point of view it is not only an interesting academic example. In fact, it may serve as a first building block for constructions of the many-body ground states of larger number of interacting particles. For example, it can be used as an input for the variational Jastrow-like ansatz based on analytical solutions of two-body problems \cite{Jastrow}.

\section*{Acknowledgements}
The authors would like to thank M. Gajda, M. Lewenstein, D. P{\k e}cak, and M. P\l odzie\'n for discussions and very fruitful suggestions. We also thank F. Deuretzbacher for bringing our attention to his PhD thesis. This work was supported by the (Polish) National Science Center Grant No. 2016/22/E/ST2/00555 (TS).

\section*{Author contributions statement}
P.K. and T.S. equally contributed in all stages of the project.

\section*{Additional information}

\textbf{Competing financial interests} The authors declare no competing financial interests.

\end{document}